\documentclass{article}
\usepackage{spconf,amsmath,graphicx}
\usepackage{xcolor}
\usepackage{amsfonts}
\usepackage[normalem]{ulem}
\usepackage[]{hyperref}
\usepackage{enumitem}
\usepackage{multirow}
\usepackage{amssymb}
\usepackage{graphicx}

\definecolor{ForestGreen}{RGB}{34,139,34}

\title{Deep Video Inpainting Guided by Audio-Visual Self-Supervision}
\name{Kyuyeon Kim, Junsik Jung\textsuperscript{*}\thanks{\textsuperscript{*} Equal contributions}, Woo Jae Kim\textsuperscript{*}, Sung-Eui Yoon
}%
\address{School of Computing, Korea Advanced Institute of Science and Technology (KAIST)}
\begin{document}
\ninept
\maketitle
%
\begin{abstract}
Humans can easily imagine a scene from auditory information based on their prior knowledge of audio-visual events. In this paper, we mimic this innate human ability in deep learning models to improve the quality of video inpainting. To implement the prior knowledge, we ﬁrst train the audio-visual network, which learns the correspondence between auditory and visual information. Then, the audio-visual network is employed as a guider that conveys the prior knowledge of audio-visual correspondence to the video inpainting network. This prior knowledge is transferred through our proposed two novel losses: audio-visual attention loss and audio-visual pseudo-class consistency loss. These two losses further improve the performance of the video inpainting by encouraging the inpainting result to have a high correspondence to its synchronized audio. Experimental results demonstrate that our proposed method can restore a wider domain of video scenes and is particularly effective when the sounding object in the scene is partially blinded.
\end{abstract}
\section{Introduction}
\label{sec:intro}

Imagine hearing the sound of a bird singing.
You may come up with an image of a bird flying in the sky or sitting on top of a tree.
In this fashion, humans can easily visualize a scene related to incoming auditory signals~\cite{arandjelovic2017look}.
This natural behavior is empowered by the prior knowledge of semantic mapping between the visual and auditory modalities learned from ubiquitous audio-visual events around us.
This ability to connect the dots between two modalities allows humans to restore videos better whose spatial information is corrupted. 
In other words, even though the video is partially blinded, humans can easily imagine what is happening in missing parts by listening to the corresponding audio.
Based on this intuition, our work tries to mimic this human ability in deep learning models to better solve the following video inpainting problem: filling in missing visual regions in a video, guided by the audio signal.
Hence, our goal can be articulated into answering the following question: can machines also learn to restore the visual content of a video by hearing its corresponding sound?

To achieve such goal, we exploit the audio-visual correspondence learned by the \textbf{audio-visual network (AV-Net)}~\cite{arandjelovic2018objects} to train the \textbf{video inpainting network (VI-Net)}.
The AV-Net learns to generate an audio-visual attention map that highlights visual regions which are corresponding to the synchronized audio and to capture the pseudo-class of each modality within the audio-visual pair.
In this manner, the AV-Net learns the semantic relationship within the audio-visual pairs without video labels in a self-supervised manner, without labeled videos.
There have been previous attempts to use this audio-visual correspondence for several of their unique downstream tasks, such as sounding object localization~\cite{arandjelovic2018objects, chen2021localizing} and sound source separation~\cite{zhao2018sound, ephrat2018looking}.
Unlike these attempts, we aim to leverage the prior knowledge of audio-visual correspondence for the video inpainting task, which has not been explored yet.

\begin{figure}[tbp]
  \centering
  \includegraphics[scale=0.7]{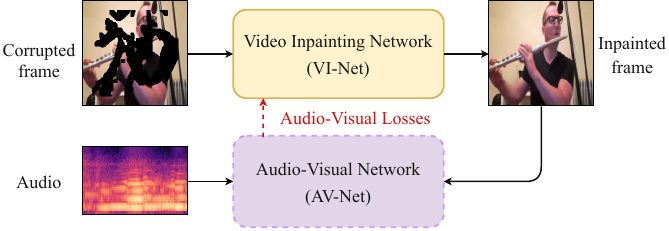}
  \vspace{-0.5em}
  \caption{Overview of our proposed method. We use the audio-visual network (AV-Net) as a guider of the video inpainting network (VI-Net) that conveys the prior knowledge of audio-visual correspondence through our proposed audio-visual losses.}
  \vspace{-1.0em}
  \label{fig:intro}
\end{figure}

As shown in Fig. \ref{fig:intro}, the AV-Net guides the VI-Net to use the corresponding audio signal as an important cue for restoring the corrupted frame.
Given the prior information of audio-visual correlation that AV-Net provides, we propose two novel audio-visual losses to convey the prior knowledge to the VI-Net: \textbf{audio-visual attention loss} and \textbf{audio-visual pseudo-class consistency loss}.
Audio-visual attention loss encourages the VI-Net to minimize the disparity of the audio-visual attention maps between the original and the inpainted frame. By doing so, the VI-Net solely focuses on restoring areas corresponding to the sounding object, making the inpainting result semantically more accurate. 
Audio-visual pseudo-class consistency loss is designed to indicate that visual and audio information from the same video should belong to the identical class. 
Using auxiliary classifiers, we encourage the VI-Net to learn that the visual features of inpainted frames and the synchronized audio features should belong to the same pseudo-classes. 
This audio-guided class consistency information can further enhance the video inpainting quality.

\begin{figure*}[htbp]
  \centering
  \includegraphics[scale=0.76]{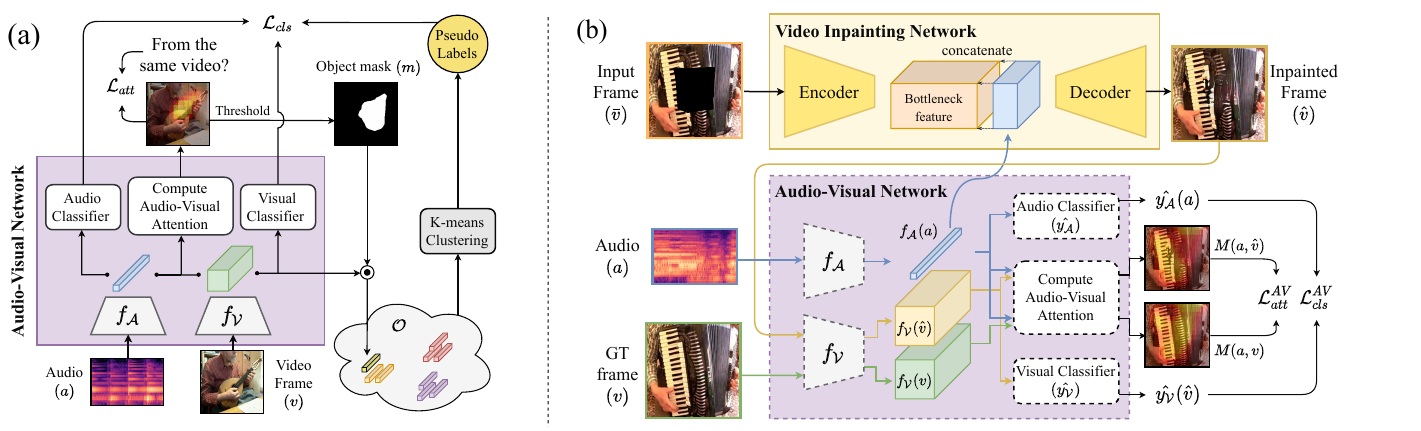}
  \vspace{-0.5em}
  \caption{Schematics of \textbf{(a) Training the audio-visual network (AV-Net)} and \textbf{(b) Training the video inpainting network (VI-Net)} with the audio-visual guidance from the pretrained AV-Net. Modules with dotted line indicate that their parameters are frozen in the training time.}
  \vspace{-1em}
  \label{fig:overall}
\end{figure*}

In summary, our main contributions are as follows:
\begin{itemize}[leftmargin=9pt,topsep=0ex,noitemsep]
    \item To enhance the video inpainting quality, we propose a novel approach that utilizes the inherent sound from the video itself.
    \item Based on the pretrained AV-Net, we propose two novel losses that enable the VI-Net to utilize the inherent sound of a video for restoring corrupted frames.
    \item Experimental results show that our approach is especially effective when restoring the frame whose sounding object in the scene is partially blinded.
\end{itemize}

\section{Related Work}
\label{sec:related_work}
In this section, we briefly discuss two research domains relevant to our work: deep video inpainting and audio-assisted visual synthesis.

\vspace{0.5ex}\noindent
\textbf{Deep video inpainting.}
Video inpainting is a challenging problem aiming to restore missing regions in consecutive frames with spatially and temporally plausible content~\cite{kim2019deep}.
Recent approaches have achieved significant improvements via deep learning by using encoder-decoder-based architectures~\cite{wang2019video}.
Along this line, there have been attempts of adopting optical flows~\cite{huang2016temporally, xu2019deep, gao2020flow}, novel architectures~\cite{chang2019learnable, oh2019onion}, attention modules~\cite{ zeng2020learning, lee2019copy}, and adversarial mechanisms \cite{chang2019free}.
Despite these successes, little attention has been given to employing the audio signal, which is the innate correspondence prior within a video.
Thus, our work takes a pioneering step to demonstrate a generic method of utilizing audio signals to the video inpainting problem.

\vspace{0.5ex}\noindent
\textbf{Audio-assisted visual synthesis.}
Audio has been used as an effective prior for synthesizing images or video frames, but in limited application domains. Such domains include audio-based adversarial image generation~\cite{wan2019towards}, talking face synthesis~\cite{jamaludin2019you, koumparoulis2020audio}, and speakers' face super-resolution~\cite{meishvili2020learning}.
Compared to these previous approaches, our work has the following distinctions.
While \cite{wan2019towards} used human-labeled videos to obtain semantic knowledge, our work utilizes an audio-visual relationship learned from the self-supervised training procedure.
We also consider a broader scope of audio-visual events occurring in the real world, rather than managing only the video of talking faces as in~\cite{jamaludin2019you,koumparoulis2020audio,meishvili2020learning}.

\section{Proposed Method}
\label{sec:method}
In this section, we explain our audio-guided video inpainting framework.
Fig. \ref{fig:overall} shows an overview of this framework consisting of two main parts: the audio-visual network (AV-Net) and the video inpainting network (VI-Net). 
We first review the AV-Net (Sec. \ref{ssec:AV-Net}). Then, we provide details on two novel losses derived from the AV-Net that are used to train the VI-Net: audio-visual attention loss and audio-visual pseudo-class consistency loss (Sec. \ref{ssec:TrainingVI-Net}).

\subsection{The audio-visual network}
\label{ssec:AV-Net}
Let $\mathcal{X} = \left\{\left(a_i, v_j\right) \mid 1 \leq i \leq N, 1 \leq j \leq N \right\}$ denote a set of audio-visual pairs such that a pair $(a_i, v_j)$ is sampled from $N$ number of videos.
Here, $a$ and $v$ each represents the audio signal and the video frame.
Given the pair $(a_i, v_j)\in\mathcal{X}$ as input, we aim to obtain the prior information in two forms: audio-visual attention map and pseudo-class of each input's modality.
The former considers a pair $(a_i, v_j)$ where $a_i$ and $v_j$ are each randomly sampled from $N$ videos, while the latter considers only a pair drawn from the same video (i.e., $i = j$).
Our training methodology of the AV-Net refers to \cite{arandjelovic2018objects, hu2020discriminative}.

\vspace{0.5ex}\noindent
\textbf{Audio-visual attention map.}
As shown in Fig.~\ref{fig:overall}~(a), the AV-Net consists of two convolutional sub-networks for feature extraction: audio network $f_\mathcal{A}$ and visual network $f_\mathcal{V}$.
From two sub-networks, we extract an audio feature $f_\mathcal{A}(a_i)\in\mathbb{R}^{c}$ and a visual feature map $f_\mathcal{V}(v_j)\in\mathbb{R}^{h\times w\times c}$.
Note that $h\times w$ and $c$ denote spatial and channel dimensions, respectively.
Then, we obtain the similarity map of $\mathbb{R}^{h\times w}$ by computing the scalar product between $L_2$-normalized $f_\mathcal{A}(a_i)$ and $f_\mathcal{V}(v_j)$ along the channel dimension for each of the spatial units within $f_\mathcal{V}(v_j)$.
The similarity map then describes how strongly each spatial location of $f_\mathcal{V}(v_j)$ reacts to the audio descriptor $f_\mathcal{A}(a_i)$.
Finally, we apply a sigmoid operation to this similarity map to obtain the audio-visual attention map $M(a_i, v_j)\in\mathbb{R}^{h\times w}$. 

Intuitively, the attention map $M(a_i, v_j)$ would show high attention in the area that semantically corresponds to both the given audio $a_i$ and the video frame $v_j$. Based on this intuition, the objective of training the AV-Net can be formulated into a binary classification problem as follows:
\begin{equation}
\label{eq:L_att}
    \mathcal{L}_{att} = \texttt{BCE}\left(y_{corr}, \texttt{GMP}(M(a_i,v_j))\right),
\end{equation}
where $\texttt{BCE}(\cdot, \cdot)$ denotes the binary cross-entropy loss, $\texttt{GMP}(\cdot)$ denotes the global max-pooling operation, and $y_{corr}$ denotes a binary label that indicates whether the audio-visual pair comes from the same video.
By minimizing the cross-entropy between $y_{corr}$ and the largest value of the attention map $M(a_i, v_j)$, the network is encouraged to maximize the attention values in regions that correspond to the given audio $a_i$, and to suppress them when audio-visual pairs do not match.

\vspace{0.5ex}\noindent
\textbf{Pseudo-class prediction.}
Pseudo-labels of audio and visual features are also used to stabilize the training of the AV-Net.
For pseudo-label extraction, we use matching audio-visual pairs $(a_i, v_i)\in\mathcal{X}$.
We apply set threshold to the attention map $M(a_i, v_i)$ to obtain a binary mask $m_i\in \left\{0, 1\right\}^{h\times w}$.
Using this attention-based binary mask, we compute the object representation $o_i\in\mathbb{R}^{c}$ from the visual feature $f_\mathcal{V}(v_i)$ to pick out the area where the audio-visual event is present.
In specific, $o_i = \texttt{GAP}\left(m_i \odot f_\mathcal{V}(v_i)\right)$, where $\texttt{GAP}(\cdot )$ and $\odot$ denote the global average pooling operation and the channel-wise Hadamard product, respectively.
We finally perform a K-means clustering on the set of object descriptors $\mathcal{O}=\left\{o_1,o_2,\cdots,o_N\right\}$ to assign each of them a pseudo-label corresponding to the cluster to which it belongs.

With these pseudo-labels set as ground truth, the network and classifiers are trained by minimizing the following classification objective:
\begin{equation}
\label{eq:L_cls}
    \mathcal{L}_{cls} = \texttt{CE}\left(y_p(o_i), \hat{y_\mathcal{A}}(a_i)\right) +  \texttt{CE}\left(y_p(o_i), \hat{y_\mathcal{V}}(v_i)\right),
\end{equation}
where $\texttt{CE}(\cdot, \cdot)$ denotes the categorical cross-entropy loss between two logit vectors and $y_p(o_i)$ represents the one-hot pseudo-label of $o_i$. $\hat{y_\mathcal{A}}(a_i)$ and $\hat{y_\mathcal{V}}(v_i)$ indicate the logit vectors from the linear classifiers $\hat{y_\mathcal{A}}$ and $\hat{y_\mathcal{V}}$ given $a_i$ and $v_i$, respectively.

We train the AV-Net with Eq.~\ref{eq:L_att} and Eq.~\ref{eq:L_cls} in an alternate manner, as two objectives mutually improves the overall performance~\cite{zhou2016learning}.

\subsection{Training the video inpainting network}
\label{ssec:TrainingVI-Net}
In this sub-section, we describe our two novel losses derived from the AV-Net to further improve the training of the VI-Net.
Suppose a pair of a corrupted video frame $\bar{v}$ and its ground truth frame $v$.
Then, the VI-Net returns an inpainted frame $\hat{v}$ given the corrupted frame $\bar{v}$.

\vspace{0.5ex}\noindent
\textbf{Audio-visual attention loss.}
We exploit the ability of the AV-Net to localize the sounding object in order to design our novel audio-visual attention loss.
The audio-visual network takes video frame $v$ and its paired audio $a$ as inputs and generates an attention map $M(a, v)$ which highlights the area matching the given audio $a$.
In the same way, the attention map $M(a, \hat{v})$ can be obtained by replacing $v$ with $\hat{v}$.
The key idea is that if the spatial content of the audio-visual event are successfully recovered in $\hat{v}$, the attention maps $M(a, \hat{v})$ and $M(a, v)$ should be identical. Otherwise, $M(a, \hat{v})$ would be vastly different from $M(a, v)$, especially in the area where the audio-visual event takes place.

From the investigation above, we observe that minimizing the difference between these two attention maps would reduce the disparity between $v$ and $\hat{v}$.
Hence, we propose the following audio-visual attention loss:
\begin{equation}
    \mathcal{L}^{AV}_{att}=
    \frac{1}{hw}\left\lVert M(a, v) - M(a, \hat{v})\right\rVert^2_2,
\end{equation}
where $h$ and $w$ are the height and width of $M(\cdot, \cdot)$, respectively.
This objective encourages the VI-Net to reconstruct the corrupted frame in a way such that the audio-visual attention map of the inpainted frame $M(a, \hat{v})$ is similar to the attention map of the ground truth frame $M(a, v)$.
As a result, the inpainting network can better restore the missing part within sound-salient areas by filling it with content or texture that actively reacts to the given audio feature.
This property cannot be found in common reconstruction losses (e.g., $L_1$ loss), which ignore additional cues from the audio.

\vspace{0.5ex}\noindent
\textbf{Audio-visual pseudo-class consistency loss.}
To further improve the performance of the VI-Net, we additionally guide it with the class-consistency information between the audio and video frame inputs. 
The audio and visual information from a synchronized video should semantically belong to the same class.
Hence, by learning that the restored frame $\hat{v}$ should belong to the same class as the corresponding audio $a$, the VI-Net could better reconstruct $\hat{v}$ such that it is more similar to the ground truth frame $v$.

We inject the audio information to the VI-Net by concatenating the audio feature $f_\mathcal{A}(a)$ to the bottleneck feature from the encoder of the VI-Net (the upper part of Fig.~\ref{fig:overall}~(b)).
Note that we broadcast the audio feature $f_\mathcal{A}(a)$ to the spatial dimension of the bottleneck feature before the concatenation.
As the pretrained AV-Net can already predict the pseudo-class of the audio $a$, we set this as a guideline to determine whether the inpainted frame $\hat{v}$ has coherent content.
Therefore, we design the audio-visual pseudo-class consistency loss as follows:
\begin{equation}
    \mathcal{L}^{AV}_{cls}=
    \texttt{CE}\left(\hat{y_\mathcal{A}}(a),  \hat{y_\mathcal{V}}(\hat{v})\right),
\end{equation}
where $\hat{y_\mathcal{A}}(a)$ and $\hat{y_\mathcal{V}}(\hat{v})$ denote the logit vectors from the linear classifiers given $a$ and $\hat{v}$, respectively. Note that the linear classifiers are also pretrained and frozen as parts of the AV-Net.
Audio-visual pseudo-class consistency loss guides the VI-Net to synthesize a frame $\hat{v}$ that is class-consistent with the synchronized audio $a$.


\vspace{0.5ex}\noindent
\textbf{Total loss.}
To train the VI-Net, we use the final loss as follows:
\begin{equation}
\label{eq:total_loss}
    \mathcal{L}=
    \lambda_{L_1}\mathcal{L}_{L_1}+
    \lambda_{adv}\mathcal{L}_{adv}+
    \lambda^{AV}_{att}\mathcal{L}^{AV}_{att}+
    \lambda^{AV}_{cls}\mathcal{L}^{AV}_{cls},
\end{equation}
where $\mathcal{L}_{L_1}$ and $\mathcal{L}_{adv}$ respectively denote the $L_1$ loss and the adversarial loss from T-PatchGAN~\cite{chang2019free}. Note that these two losses are borrowed from~\cite{zeng2020learning}, which is our baseline VI-Net.
The VI-Net is optimized jointly with our proposed losses $\mathcal{L}^{AV}_{att}$ and $\mathcal{L}^{AV}_{cls}$.
Hence, the network learns to consider the audio-visual consistency while minimizing the visual difference.
The weights for each loss are empirically set as follows: $\lambda_{L_1}=1$, $\lambda_{adv}=0.01$, $\lambda^{AV}_{att}=2$, and $\lambda^{AV}_{cls}=1$.

\begin{figure*}[tbp]
  \centering
  \includegraphics[scale=0.51]{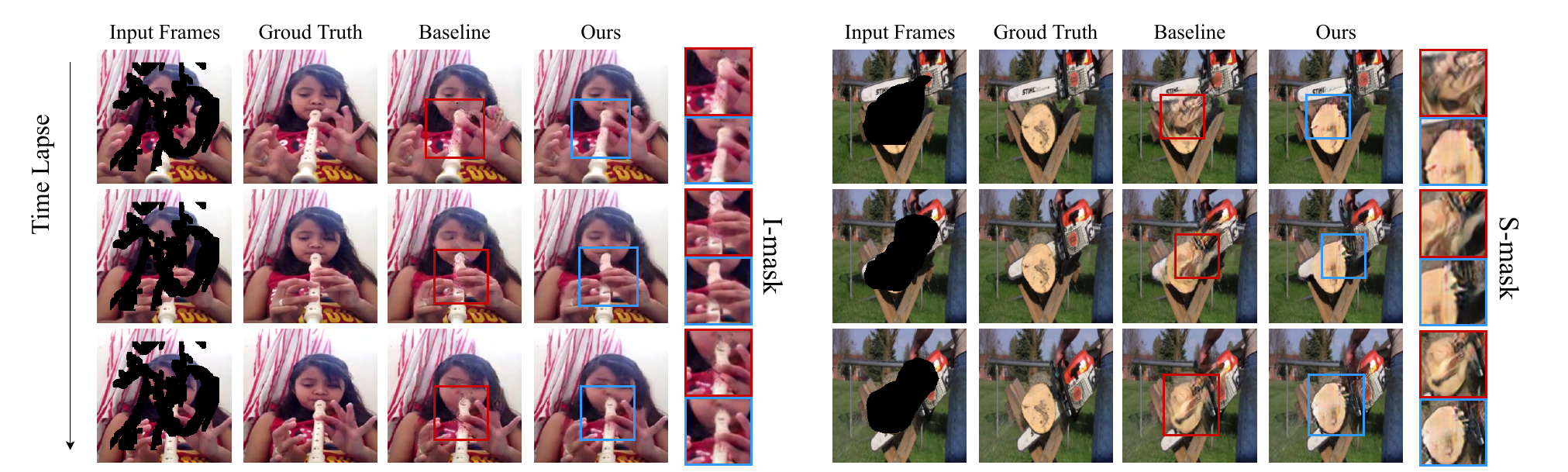}
  \vspace{-1em}
  \caption{Qualitative results of two samples from AVE dataset blinded by \textbf{I-masks} (left) and \textbf{S-masks} (right). While the baseline STTN without audio signals shows more artifacts around the object and produces blurry result, our method produces more realistic and clearer results.}
  \vspace{-1em}
  \label{fig:res}
\end{figure*}

\section{Experiments}
\label{sec:experiments}

\subsection{Experimental settings}
\label{subsec:4.1}
\vspace{0.5ex}\noindent
\textbf{Datasets.}
We adopt \textbf{AVE}~\cite{tian2018audio} and \textbf{MUSIC-Solo}~\cite{zhao2018sound} dataset to show the effectiveness of our approach.
AVE dataset contains 4113 video clips covering 29 categories of diverse real-life audio-visual events.
MUSIC-Solo dataset contains 493 video clips with 11 categories that exclusively cover solo performances of diverse musical instruments.
We follow the official split of AVE dataset. On the other hand, we randomly split MUSIC-Solo dataset into 343/50/100 for train/validation/test since there is no designated split.

Moreover, we evaluate our method on two types of maskings: \textbf{I-mask} and \textbf{S-mask}.
I-masks irregularly blind the pixels with random strokes and shapes. We adopt the subset of NVIDIA Irregular Mask Dataset~\cite{liu2018partialinpainting}.
For testing, we randomly pick three masks with a blinding ratio of 20.0\%, 27.7\%, and 28.4\%, respectively.
We also design S-masks to blind the region which corresponds to the sounding object.
We collect S-masks by eroding the object mask $m_i$ mentioned in Sec.~\ref{ssec:AV-Net} until the spatial area of the masking covers 20\% of the image.
This ratio refers to the approximate proportion of the region that the sounding object occupies in the video scene.

\vspace{0.5ex}\noindent
\textbf{Preprocessing.}
Given a video clip, we extract video frames at 8 fps and resample its mono-channel audio at 16 kHz.
Then, the video frame is resized to the spatial size of 256$\,\times\,$256 and then randomly cropped (for training) or resized (for testing) into 224$\,\times\,$224.
The audio is sampled by retrieving a 1-second segment, and converted to the log-scale mel spectrogram with 0.01-second window size, half-window hop length, and 80 mel bins, finally treated as a single-channel matrix with the spatial dimension of 201$\,\times\,$80.

\vspace{0.5ex}\noindent
\textbf{Audio-visual network} We follow~\cite{arandjelovic2018objects, hu2020discriminative} to implement the audio-visual network (AV-Net). For visual and audio sub-networks $f_\mathcal{V}$ and $f_\mathcal{A}$, we use ResNet-18-based architectures as in~\cite{hu2020discriminative}.

\vspace{0.5ex}\noindent
\textbf{Video inpainting baseline.}
We adopt one of the state-of-the-art architectures, the Spatial-Temporal Transformer Network (STTN)~\cite{zeng2020learning} as our baseline model.
As our major interest lies in inpainting videos with audio-visual events, our choice of video dataset is different from the original work~\cite{zeng2020learning}.
Therefore, we train the STTN on the aforementioned datasets from scratch, without audio signals.

\vspace{0.5ex}\noindent
\textbf{Training details.}
To train the AV-Net, we adopt Adam optimizer with the learning rate of \texttt{5e-5} for AVE and \texttt{1e-4} for MUSIC-Solo dataset.
The batch size is set to 32 for both datasets.
Furthermore, we set the threshold value to 0.07 while obtaining binary masks and the number of clusters to 10 while collecting the pseudo-classes of object representations.
While training the AV-Net for 4 epochs total, the learning rate is decayed by 0.1 after 2 epochs.
Then, the parameters of the pretrained AV-Net are frozen while training the VI-Net.
For AVE dataset, we train the VI-Net using Adam optimizer with the initial learning rate of \texttt{1e-4} decayed by 0.1 for every 100k iterations for a total of 350k iterations.
For MUSIC-Solo dataset, due to the lack of training data, we fine-tune the VI-Net pretrained on AVE dataset using Adam optimizer for a total of 100k iterations with the learning rate of \texttt{1e-5} for first 50k iterations, and \texttt{1e-6} for the remaining iterations.
For both datasets, the batch size is set to 8.

\vspace{0.5ex}\noindent
\textbf{Evaluation metrics.} 
The quantitative result is reported using three widely-used metrics: PSNR~\cite{xu2019deep}, SSIM~\cite{wang2004image}, and video-based Fréchet Inception Distance (VFID)~\cite{chang2019free}.
In detail, PSNR and SSIM are standard metrics to assess the synthesized scenes, whereas VFID quantifies the perceptual difference compared to the ground truth.
\renewcommand{\tabcolsep}{3pt}
\renewcommand{\arraystretch}{1.1}
\begin{table}[htbp]
\centering
\resizebox{0.48\textwidth}{!}{
\begin{tabular}{|c|c|c||c|c|c||c|c|c|}
\noalign{\smallskip}\hline
\multicolumn{3}{|c||}{Method} & \multicolumn{3}{c||}{I-mask} & \multicolumn{3}{c|}{S-mask} \\
\cline{1-9}
  & $\scriptstyle\mathcal{L}^{AV}_{att}$ & $\scriptstyle\mathcal{L}^{AV}_{cls}$ & PSNR$\uparrow$ & SSIM$\uparrow$ & VFID$\downarrow$ & PSNR$\uparrow$ & SSIM$\uparrow$ & VFID$\downarrow$ \\
\hline\hline
 Baseline & - & - & 30.76 & 93.45 & 3.549 & 26.58 & 91.93 & 5.553 \\
 w/ Ours & - & \checkmark & 30.81 & 93.55 & 3.356 & 26.83 & 92.21 & 5.305 \\
 w/ Ours & \checkmark & - & 30.94 & 93.61 & 3.273 & 27.16 & 92.47 &  5.271 \\
 w/ Ours & \checkmark & \checkmark & \textbf{31.18} & \textbf{93.65} & \textbf{3.184} & \textbf{27.32} & \textbf{92.69} & \textbf{4.961} \\
\hline
\end{tabular}
}
\caption{Quantative evaluation and ablation study of applying our method on \textbf{AVE} dataset with two different types of masks. $\uparrow$ indicates that higher is better and $\downarrow$ means that lower is better.}
\vspace{-1em}
\label{table:ave}
\end{table}

\renewcommand{\tabcolsep}{3pt}
\renewcommand{\arraystretch}{1.1}
\begin{table}[htbp]
\centering
\resizebox{0.48\textwidth}{!}{
\begin{tabular}{|c|c|c||c|c|c||c|c|c|}
\noalign{\smallskip}\hline
\multicolumn{3}{|c||}{Method} & \multicolumn{3}{c||}{I-mask} & \multicolumn{3}{c|}{S-mask} \\
\cline{1-9}
  & $\scriptstyle\mathcal{L}^{AV}_{att}$ & $\scriptstyle\mathcal{L}^{AV}_{cls}$ & PSNR$\uparrow$ & SSIM$\uparrow$ & VFID$\downarrow$ & PSNR$\uparrow$ & SSIM$\uparrow$ & VFID$\downarrow$ \\
\hline\hline
 Baseline & - & - & 29.49 & 93.85 & 4.316 & 26.47 & 91.88 & 5.706 \\
 w/ Ours & - & \checkmark & 29.60 & 93.84 & 4.191 & 26.95 & 92.38 & 5.205 \\
 w/ Ours & \checkmark & - & 29.66 & \textbf{93.87} & 4.221 & 27.05 & 92.40 & 5.194 \\
 w/ Ours & \checkmark & \checkmark & \textbf{29.68} & 93.84 & \textbf{4.092} & \textbf{27.12} & \textbf{92.53} & \textbf{4.929} \\
\hline
\end{tabular}
}
\caption{Quantative evaluation and ablation study of applying our method on \textbf{MUSIC-Solo} dataset with two different types of masks. $\uparrow$ indicates that higher is better and $\downarrow$ means that lower is better.}
\vspace{-1em}
\label{table:music}
\end{table}

\subsection{Result and discussion}
\vspace{0.5ex}\noindent
We test our method on 4 different experimental setups derived from the combinations of video and mask datasets mentioned in Sec. \ref{subsec:4.1}.
Table \ref{table:ave} shows that adopting our proposed audio-visual objectives outperforms the visual-only baseline on AVE dataset for all suggested metrics.
As shown in Table \ref{table:music}, our method also performs substantially well on MUSIC-Solo dataset with video scenes strictly related to musical instruments.
Performance improvements over two different video datasets also show that our method is effective not only in domain-specific videos such as MUSIC-Solo dataset but also in videos with a broader domain such as AVE dataset.
Ablation studies in Table \ref{table:ave} and \ref{table:music} imply that our two losses harmoniously give a positive impact on the inpainting quality, with the audio-visual attention loss showing a bigger influence.

One interesting point is that performance gains on S-masks are larger than those on I-masks.
As shown in Table \ref{table:ave}, on AVE dataset masked with I-masks, our method of applying both audio-visual losses improves the baseline PSNR and VFID by 0.42 and 0.365, respectively.
On the same dataset with S-masks, our method shows a larger PSNR increase of 0.74 and VFID improvement of 0.592.
The same tendency is shown in Table \ref{table:music} on the MUSIC-Solo dataset.
In the case of I-masks, PSNR and VFID improvement each shows 0.19 and 0.224 compared to the baseline.
On the other hand, improvements are greater in the case of S-masks, showing PSNR increase of 0.65 and VFID improvement of 0.777.
Recalling that S-masks are designed to mask audio-visual events, this tendency indicates that our method indeed effectively restores those regions.
This shows that the audio-visual correspondence given as the prior information allows the video inpainting model to better restore regions corresponding to audio-visual events.

Fig. \ref{fig:res} demonstrates that our method produces more pleasing results for both types of masking.
While the baseline model produces blurry artifacts around the sounding object, our approach can synthesize plausible results.
Particularly, when the audio-visual event is partially deteriorated (by S-masks), the baseline fails to generate a realistic scene in the blinded area.
In contrast, our method successfully restores the frame with clearer and comprehensible content while preserving the audio-visual coherency.

\section{Conclusion}
\label{sec:conclusion}
In this paper, we investigate a novel approach to using audio for video inpainting tasks by employing audio-visual self-supervision.
We adopt the audio-visual network to bridge the gap between visual and audio modality, then use its functionalities to further guide the video inpainting network through proposed two novel losses: audio-visual attention loss and audio-visual pseudo-class consistency loss.
Experimental results on two different audio-visual datasets -- AVE and MUSIC-Solo dataset -- with two types of masking -- I-mask and S-mask -- show that our approach improves the performance of the video inpainting network compared to the baseline.



\vfill\pagebreak

\bibliographystyle{IEEEbib}
\bibliography{refs}

\end{document}